\documentclass[12pt]{article}
\usepackage[polish,english]{babel}
\usepackage[latin1]{inputenc}
\usepackage{graphicx}
\usepackage{times}
\usepackage[T1]{fontenc}
\usepackage{epsfig}

\def\lab#1{\label{eq:#1}}

\def\br{\begin{eqnarray}}
\def\er{\end{eqnarray}}
\def\be{\begin{equation}}
\def\ee{\end{equation}}
\def\({\left(}
\def\){\right)}

\def\rlx{\relax\leavevmode}
\def\IR{\rlx\hbox{\rm I\kern-.18em R}}

\def\vt{\vartheta}
\def\u2{\mid u\mid^2}

\begin{document}
\begin{titlepage}

\begin{center}
{\bf Some (3+1) dimensional vortex solutions of the $CP^N$ model}
\end{center}

\vspace{.5cm}

\begin{center}
{L. A. Ferreira~$^{\star}$, P. Klimas~$^{\star}$ and W. J. Zakrzewski~$^{\dagger}$}

\vspace{.5 in}
\small

\par \vskip .2in \noindent
$^{(\star)}$Instituto de F\'\i sica de S\~ao Carlos; IFSC/USP;\\
Universidade de S\~ao Paulo  \\ 
Caixa Postal 369, CEP 13560-970, S\~ao Carlos-SP, Brazil\\

\par \vskip .2in \noindent
$^{(\dagger)}$~Department of Mathematical Sciences,\\
 University of Durham, Durham DH1 3LE, U.K.

\end{center}

\begin{abstract}
We present a class of solutions of the $CP^N$ model in (3+1) dimensions.
We suggest that they represent vortex-like configurations. We also discuss
some of their properties. We show that some configurations of vortices
have a divergent 
energy per unit length while for the others  such an energy has a minimum
for a very special orientation of vortices. We also discuss the Noether
charge densities of these vortices.

\end{abstract}


\end{titlepage}

\section{Introduction}

Exact topological soliton solutions are rare specialy in dimensions
higher than two. Their existence depends on very special structures,
and the most well know examples are those provided by self-dual or BPS
solutions like instantons and monopoles. In this paper we present
exact vortex like solutions in ($3+1$) dimensions for the well known
$CP^N$ model \cite{dadda,wojtekbook}. The class of exact solutions we 
construct is in fact very large, and consists of arbitrary functions
of two variables, namely the combinations $x^1+i\, x^2$ and $x^3+x^0$,
where $x^{\mu}$, $\mu=0,1,2,3$ are the Cartesian coordinates of four
dimensional Minkowski space-time.  Among these solutions there is a
special subset which corresponds to straight vortex solutions with
waves traveling along them with the speed of light. Our results
correspond, in fact, to a specialization to $CP^N$ of those obtained in
\cite{CP1vortex,CPNvortex} for extensions of the Skyrme-Faddeev
model. 

One of the features we explore in detail in this paper is the
structure of the energy for parallel but separated vortices. The total
energy of such a solution is obviously infinite since it does not depend
on one of the space coordinates. However, the energy per unit of
length of the vortex is finite and is made out of two parts. The first one
is purely topological and corresponds to the energy of the ($2+0$)
dimensional $CP^N$ lumps. This part does not depend on the
separation of the vortices and is proportional to the two dimensional
topological charge. The second part involves the energy coming from
the waves and it does depend on the separation of the vortices. It
is, in fact, proportional to a combination of the Noether charges
associated with the $U(1)^N$ subgroup of $SU(N+1)$. We make a detailed
study of this part of the energy and discover that the solutions can
be put in two main classes according to the behavior of the energy for
large separation of the vortices. For one class the energy per unit of
length goes to zero for large separations and for the other it
diverges. In addition, both classes possess local minima of the
energy per unit of length for finite separations of the vortices. We
would like to stress that such a dependence of the energy  on the separation
does not lead to a force among the vortices.  The total
energy is infinite for all such solutions and only the energy per unit
of length is finite and dependent on the separation. Our analysis
shows that the densities of topological and Noether charges have
practically the same location for any given solution. In addition,
their topological charge densities are peaked at the location of vortices but the 
densities of the Noether charges are concentrated only near the location of each vortex.
 As far as we are aware such exact solutions for ($3+1$)
dimensional solutions  of the $CP^N$ are novel and may be important
in some applications.

We consider the $CP^{N}$ model in $3+1$ dimensional Minkowski
spacetime defined by the Lagrangian density 
\begin{eqnarray}\label{lagr1}
\mathcal{L}=M^2(D_{\mu}{\cal Z})^{\dagger}D^{\mu}{\cal Z}, \qquad
{\cal Z}^{\dagger}\cdot {\cal Z}=1,
\label{cpnlagrangian}
\end{eqnarray}
where $M^2$ is a constant with the dimension of mass,
${\cal Z}=({\cal Z}_1,\ldots,{\cal Z}_{N+1})\in \mathcal{C}^{N+1}$ and
it satisfies the 
constraint ${\cal Z}^{\dagger}\cdot {\cal Z}=1.$. The covariant
derivative $D_{\mu}$ 
acts on ${\cal Z}$ according to 
$$D_{\mu}{\cal Z}=\partial_{\mu}{\cal Z}-({\cal Z}^{\dagger}\cdot
\partial_{\mu}{\cal Z}){\cal Z}. 
$$
 The index $\mu$ runs over the set $\mu=\{0,1,2,3\}$. The Lagrangian
 (\ref{cpnlagrangian}) is invariant under the global transformation
 ${\cal Z}\rightarrow U\,{\cal Z}$, with $U$ being a $(N+1)\times
 (N+1)$ unitary  matrix.  One of the advantages of the
 ${\cal Z}$ parametrization is that it makes that $U(N+1)$ symmetry
 explicit \cite{dadda,wojtekbook}. However, one can also  use the
 parametrization  
\begin{eqnarray}
{\cal Z}=\frac{(1,u_1,\ldots,u_N)}{\sqrt{1+|u_1|^2+\ldots+|u_N|^2}},
\label{udef}
\end{eqnarray}
which solves the constraint ${\cal Z}^{\dagger}\cdot {\cal Z}=1$. The
$u$-field parametrization does not make the $U(N+1)$ symmetry
explicit, however they transform under the $U(N)$ subgroup as
$u\rightarrow {\tilde U}\,u$, with ${\tilde U}\in U(N)$.
In terms of $u_i$'s the Lagrangian density (\ref{lagr1}) takes the form
\begin{eqnarray}\label{lagr2}
\mathcal{L}=\frac{4M^2}{(1+u^{\dagger}\cdot
  u)^2}\left[(1+u^{\dagger}\cdot
  u)\partial^{\mu}u^{\dagger}\cdot\partial_{\mu}u-(\partial^{\mu}u^{\dagger}\cdot
  u)(u^{\dagger}\cdot \partial_{\mu}u)\right] 
\end{eqnarray}
and the Euler-Lagrange equations read
\begin{eqnarray}\label{cpn_equation}
(1+u^{\dagger}\cdot
  u)\,\partial^{\mu}\partial_{\mu}u_k-2(u^{\dagger}\cdot
  \partial^{\mu}u)\,\partial_{\mu}u_k=0 
\end{eqnarray}
for each $k=1,\ldots, N$. The dot product denotes $u^{\dagger}\cdot
u=\sum_{i=1}^Nu^*_i\,u_i$. The simplest $CP^1$ case is given just by
one function $u$: ${\cal Z}=\frac{(1,\,u)}{\sqrt{1+|u|^2}}$. 

\section{The solutions}

In what follows we shall use the notation
\begin{equation}
z\equiv x^1+i\,\varepsilon_1 x^2,\qquad {\bar z}\equiv x^1-i\,\varepsilon_1 x^2,
 \qquad y_{\pm}\equiv x^3\pm\epsilon_2\,x^0
\label{zydef}
\end{equation}
with $\varepsilon_a=\pm 1$, $a=1,2$.

It is easy to check that any set of functions $u_k$ that depend on
coordinates $x^{\mu}$ in a special way, namely  
\begin{equation}u_k=u_k(z,y_{+})
\label{gensolution}
\end{equation}
is a solution of the system of equations (\ref{cpn_equation}). The Minkowski
metric in the coordinates (\ref{zydef}) becomes $ds^2=-dz\,d{\bar
  z}-dy_{+}\,dy_{-}$. It then follows that (\ref{gensolution}) 
 satisfies simultaneously $\partial^{\mu}\partial_{\mu}u_i=0$
 and $\partial^{\mu}u_i\partial_{\mu}u_j=0$ for all $i,\,j = 1,\ldots,
 N$. Hence this class of solutions is very large. 

Amongst the very many solutions of the type (\ref{gensolution}) we
have those that describe 
 vortices with waves traveling along them. Such solutions have already
 been studied in the $CP^1$ case  in \cite{CP1vortex}, whereas the
 generalization to the case $CP^N$ has been considered in
 \cite{CPNvortex}.

 The solutions presented there are also exact solutions of an extension of the 
 Skyrme-Faddeev model for a special choice of their
 parameters. From the point of view of the present work  
what matters most is that (\ref{gensolution}) satisfy the equations of
the $CP^N$ model in (3+1) 
dimensions.

 In the notation of
(\ref{zydef}) the vortex 
solution studied in $\cite{CPNvortex}$ takes the form 
$$
u_i(z, y_{+})=z^{n_i} e^{i k_iy_{+}},
$$
where $k_i$ are arbitrary real numbers and $n_i$ are positive or negative
integer numbers.   
A slightly modified vortex-like solution would be given by 
$$
u_i(z, y_{+})=(z-a_i)^{n_i} e^{i k_iy_{+}},
$$
where $a_i$ are some constant numbers (in general complex) and, in the
most general ones, 
we would replace the expression $(z-a_i)^{n_i}$ by a rational map. 

In this paper, for simplicity  and the clarity of the interpretation,
we shall consider only some very special maps, leaving more complex
ones for future work. Thus we shall consider solutions of the form 
\begin{equation}\label{sol_cpn}
u_i(z, y_{+})=(z-\delta)^{n_i} (z+\delta)^{m_i}e^{i k_iy_{+}},
\end{equation}
where $\delta$ is a real constant and $m_i$ are integers. 

\section{The energy density of the solutions}

It is easy to check that the energy density of solutions
(\ref{gensolution}) takes the form 
\begin{eqnarray}
\mathcal{H}=\frac{8M^2}{(1+u^{\dagger} \cdot
  u)^2}\left[\partial_{{\bar z}}u^{\dagger}\cdot \Delta^2
  \cdot\partial_{z}u+\partial_{y_+}u^{\dagger}\cdot \Delta^2
  \cdot\partial_{y_+}u\right],
\label{hamiltonian}
\end{eqnarray}
where 
\be
\Delta^2_{ij}\equiv(1+u^{\dagger}
\cdot u)\delta_{ij}-u_iu_j^*.
\label{deltadef}
\ee 
The first term in (\ref{hamiltonian}), when evaluated for the solutions
of the type (\ref{gensolution}), is a total derivative, {\it i.e.} 
\be
\partial_z\,\partial_{{\bar z}}\,\ln\(1+u^{\dagger}\cdot u\)= 
\frac{\partial_{{\bar
      z}}u^{\dagger}\cdot\Delta^2\cdot\partial_zu}{\(1+u^{\dagger}\cdot
  u\)^2} =\frac{\mid \Delta\cdot\partial_zu\mid^2}{\(1+u^{\dagger}\cdot
  u\)^2}.
\label{niceidentity}
\ee
When integrated over the plane $x^1\,x^2$ the first term in
(\ref{hamiltonian}) becomes proportional to the topological charge of
the vortex solution as we explain below. 

The second term in (\ref{hamiltonian}) is related, for some special
solutions of the type (\ref{gensolution}), to some
Noether charges of the $CP^N$ model. To see this we note that the $CP^N$
Lagrangian (\ref{cpnlagrangian}) is invariant under global $U(N+1)$
transformations. However, the parametrization of the fields in terms
of the $u$ fields given by (\ref{udef}), gauge fixes this symmetry to
$SU(N)\otimes U(1)$. We are interested in the subgroup $U(1)^N$ of
$SU(N)\otimes U(1)$ under which the fields $u$ transform as 
$$
u_i\rightarrow e^{i\,\alpha_i}\,u_i \qquad\qquad i=1,2,\ldots N.
$$
The Noether currents associated with these symmetries are given by
\br
J_{\mu}^{(i)}=-\frac{4\,i\,M^2}{\vt^4}\, \sum_{j=1}^N\left[
u_i^*\, \(\Delta^2\)_{ij}\,\partial_{\mu}u_j -
\partial_{\mu}u_j^*\, \(\Delta^2\)_{ji}\,u_i\right].
\lab{u1currents}
\er
If one now considers solutions of the class (\ref{gensolution}) of the
form 
\be
u_i= v_i\(z\)\,e^{i\,k_i\,y_{+}}
\label{nicevortices}
\ee
with  $k_i$ being the inverse of a wavelength, then one can show that 
\be
\frac{8M^2}{(1+u^{\dagger} \cdot
  u)^2}\;\partial_{y_+}u^{\dagger}\cdot \Delta^2
  \cdot\partial_{y_+}u = \varepsilon_2\,\sum_{i=1}^N k_i\,{J}_{0}^{(i)},
\ee
where $\varepsilon_2$ was introduced in (\ref{zydef}). 

Therefore, the
energy per unit of length of our vortices solutions of the type
(\ref{nicevortices}) has the form
\be
{\cal E}= \int dx^2\,\,\mathcal{H}=  8\pi\, M^2\, Q_{\rm Top.} + 
\varepsilon_2\,\sum_{i=1}^N k_i\,Q_{\rm Noether}^{(i)},
\label{energytopnoether}
\ee
where $Q_{\rm Top.}$ is the topological charge
and 
\be
Q_{\rm Noether}^{(i)}\equiv \int dx^1\,dx^2\, {J}_{0}^{(i)}.
\ee

 To discuss the topological properties of 
(\ref{energytopnoether}) we split it into two parts; $\mathcal{H}^{(1)}$  and
$\mathcal{H}^{(2)}$ 
corresponding to its two terms.  For a solution of
the type (\ref{sol_cpn}) the  
first term reduces to
\begin{eqnarray}
\mathcal{H}^{(1)}=\frac{8M^2}{(1+u^{\dagger}\cdot u)^2}
\left[\sum_i\psi_{ii}|u_i|^2+\sum_{i,j}(\psi_{ii}-\psi_{ij})|u_i|^2|u_j|^2\right]  
\end{eqnarray}
where, if we define  $w_{\delta}\equiv z- \delta$ and
$w_{-\delta}\equiv z+ \delta$, the functions $\psi_{ij}$ are given by  
\begin{eqnarray}
\psi_{ij}(z,\bar{z})\equiv
\frac{n_in_j}{|w_{\delta}|^2}+\frac{m_im_j}{|w_{-\delta}|^2}+
(n_im_j+n_jm_i)\frac{\bar{w}_{\delta}w_{-\delta}+w_{\delta}\bar{w}_{-\delta}}{2|w_{\delta}|^2|w_{-\delta}|^2}.     
\end{eqnarray}
The second contribution $\mathcal{H}^{(2)}$ is given by
\begin{eqnarray}
\mathcal{H}^{(2)}=\frac{8M^2}{(1+u^{\dagger}\cdot
  u)^2}\left[\sum_i
  k_i^2|u_i|^2+\sum_{i<j}(k_i-k_j)^2|u_i|^2|u_j|^2\right] 
\end{eqnarray}
and the total energy density is given by
$\mathcal{H}=\mathcal{H}^{(1)}+\mathcal{H}^{(2)}$.

\begin{figure}
\begin{center}
\includegraphics[width=0.45\textwidth]{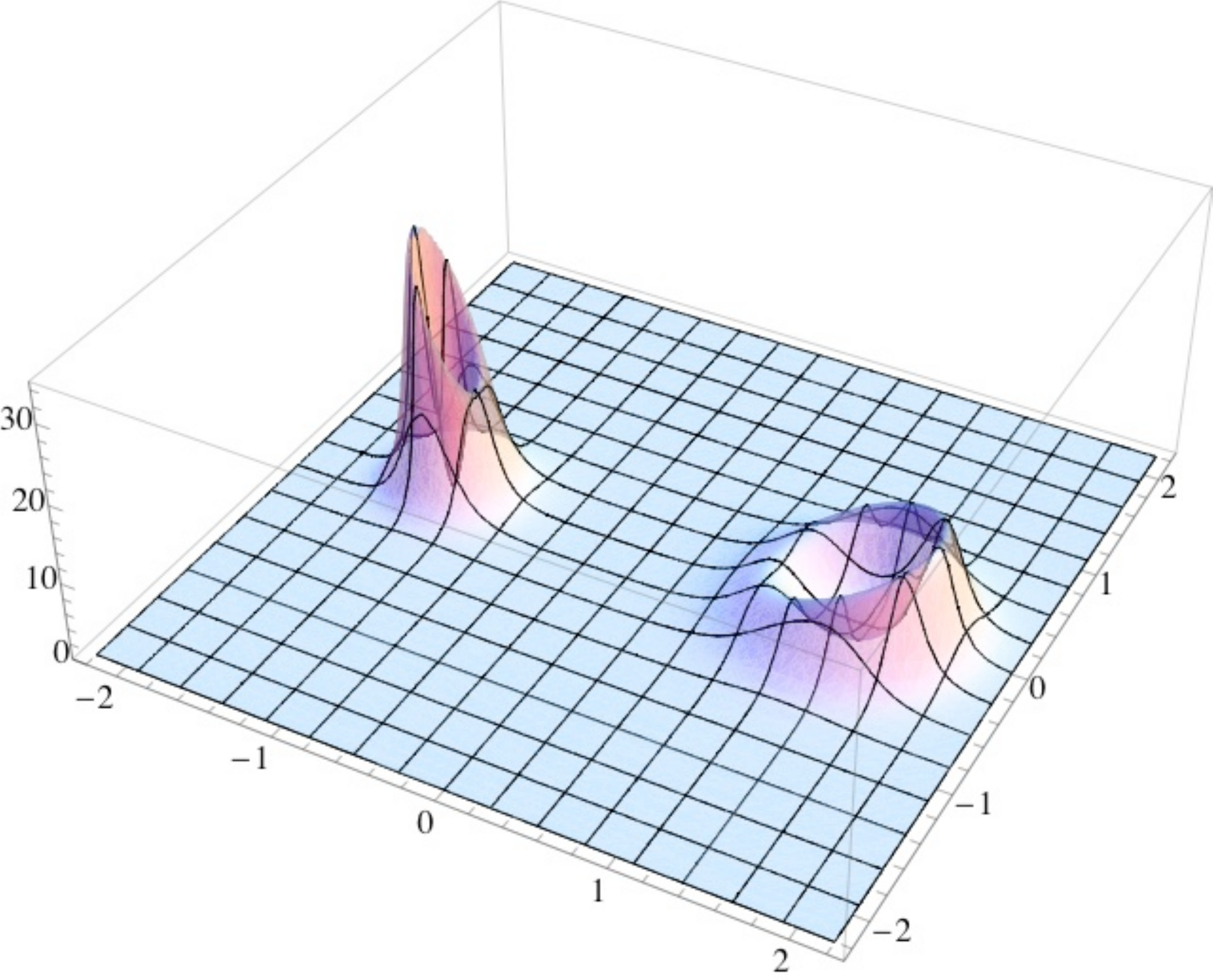}
\includegraphics[width=0.45\textwidth]{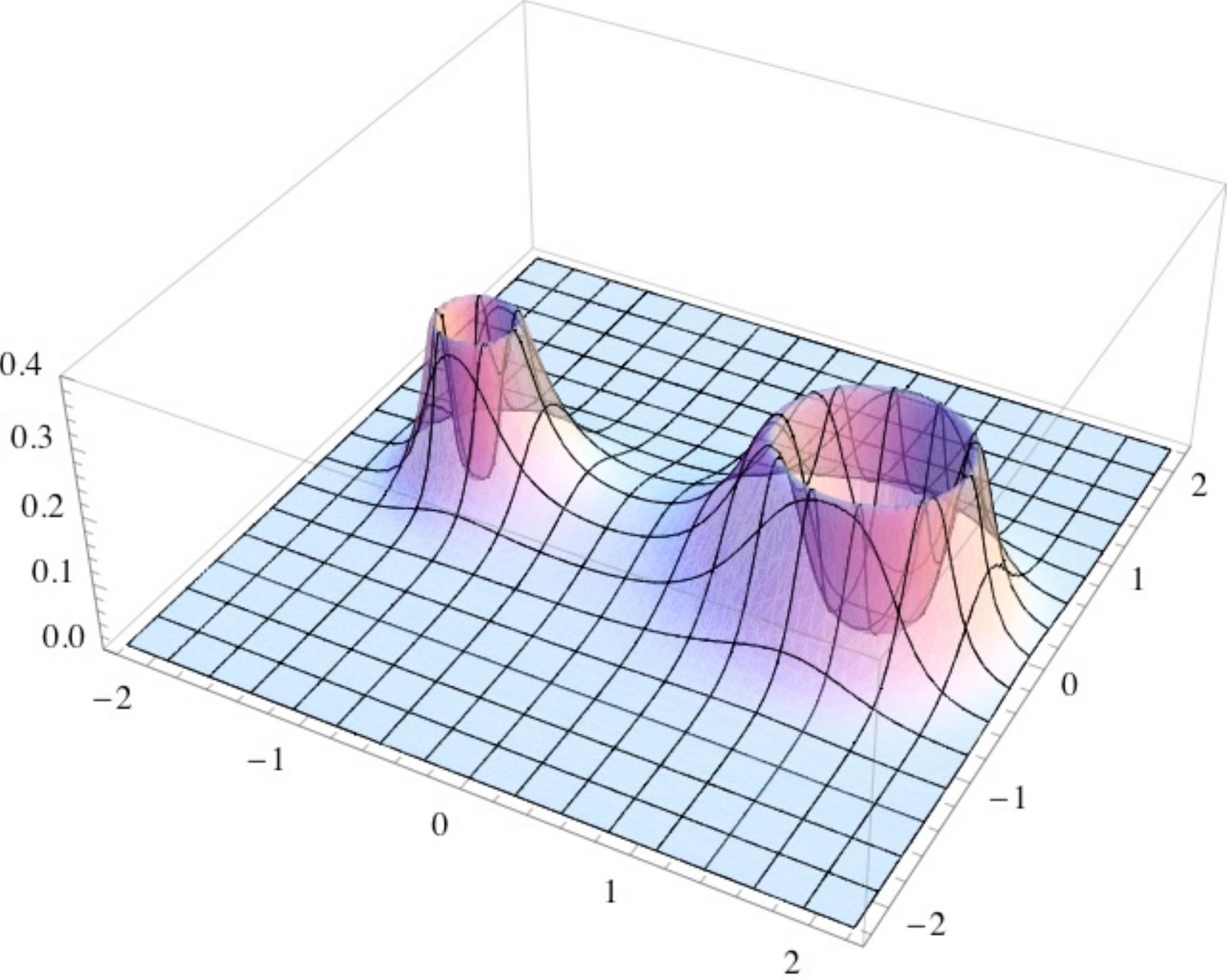} 
\caption{The energy density contributions
  $\frac{|u|^2\psi(x^1,x^2)}{(1+|u|^2)^2}$ (on left) and
  $\frac{|u|^2}{(1+|u|^2)^2}$ (on right) for the case $nm>0$. Here
  $n=3$ and $m=2$. The separation parameter $\delta=1.3$.}
\label{pos} 
\end{center}
\end{figure}

\begin{figure}
\begin{center}
\includegraphics[width=0.45\textwidth]{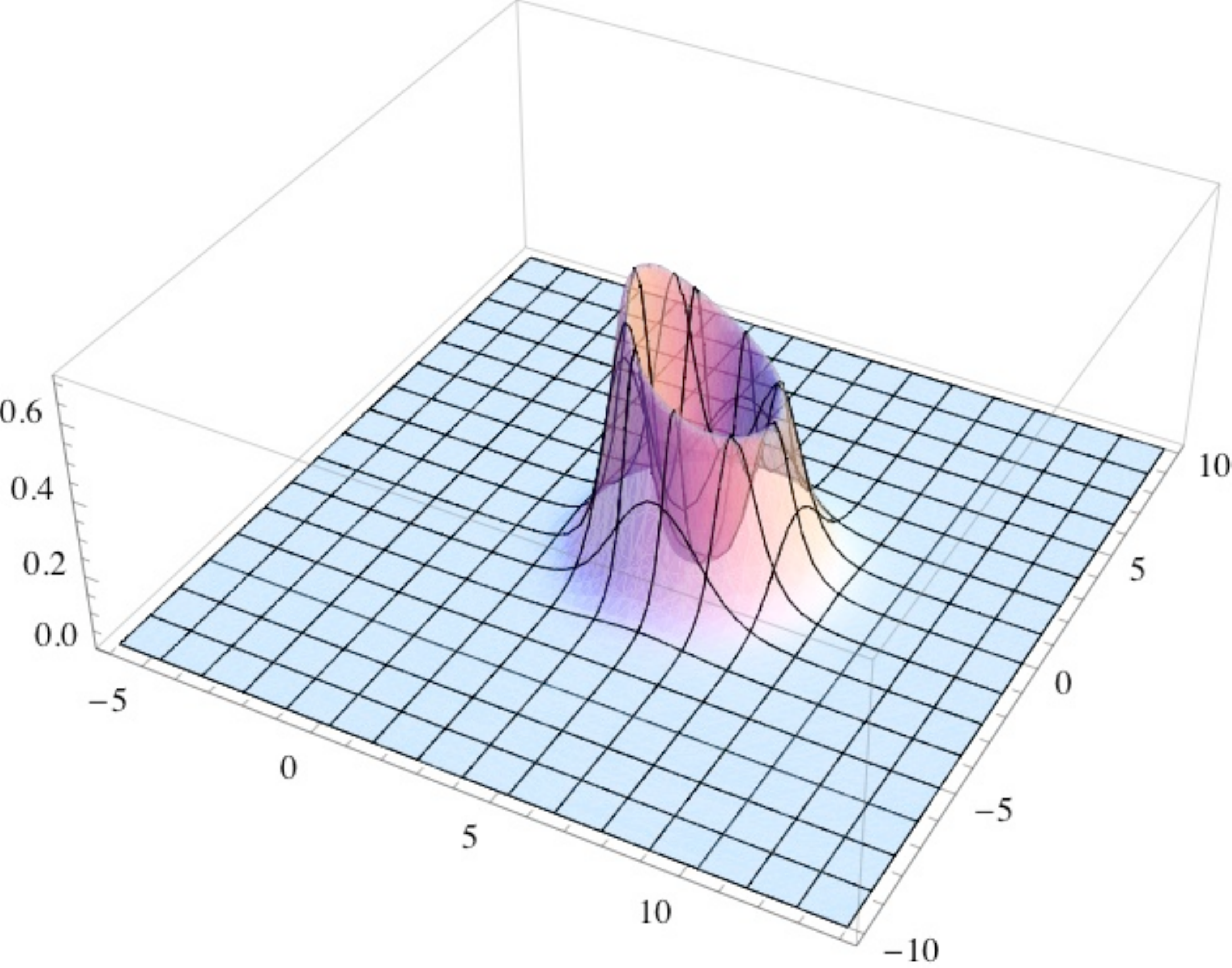}
\includegraphics[width=0.45\textwidth]{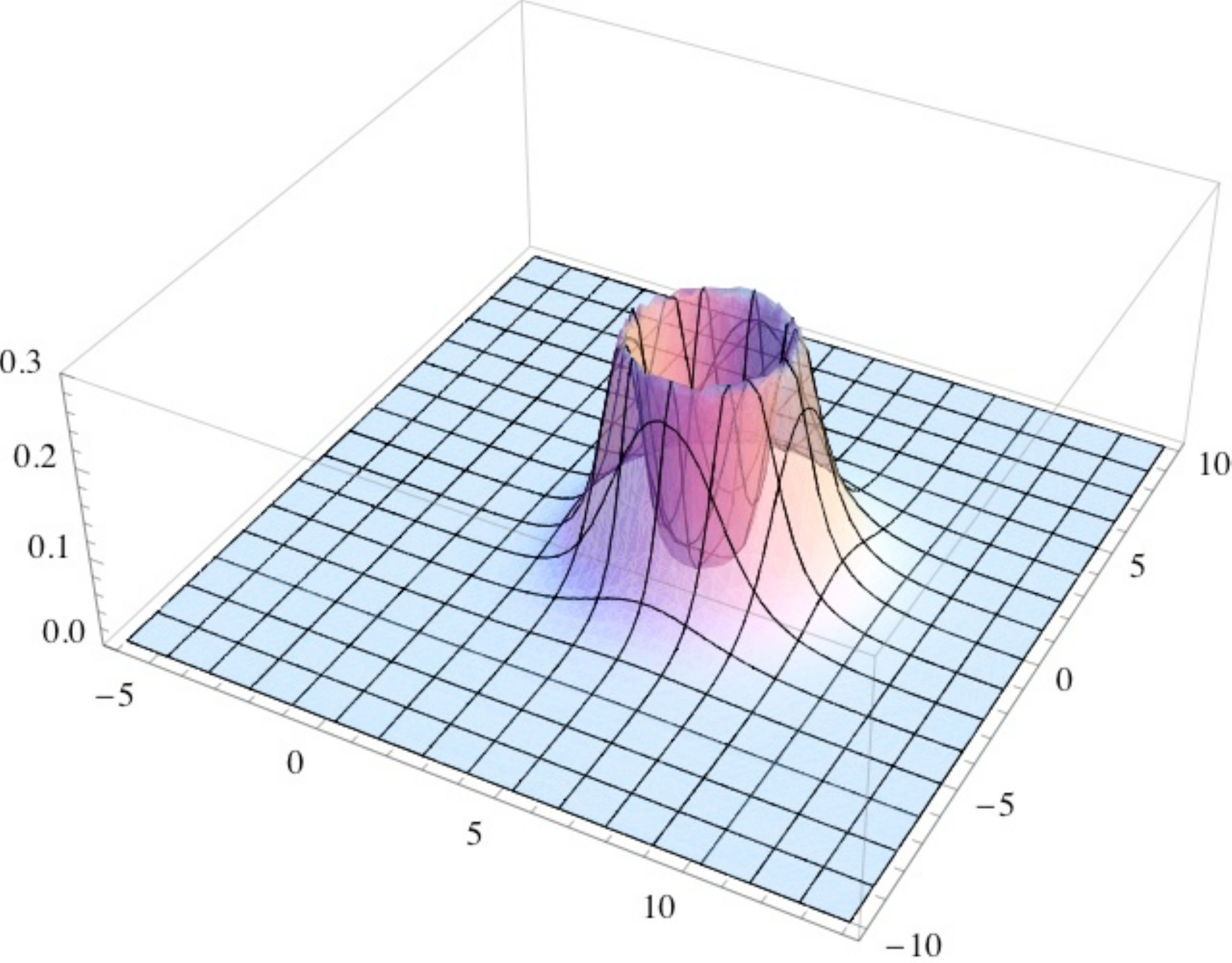}
\caption{The energy density contributions
  $\frac{|u|^2\psi(x^1,x^2)}{(1+|u|^2)^2}$ (on left) and
  $\frac{|u|^2}{(1+|u|^2)^2}$ (on right) for the case $nm<0$. Here
  $n=3$ and $m=-1$. The separation parameter $\delta=5.0$.}
\label{neg}
\end{center}
\end{figure}

In the  $CP^1$ case {\it i.e.} for 
\begin{equation} u(z,y_+)=(z-\delta)^n(z+\delta)^m e^{i k
    y_+}\label{one}\end{equation} 
the energy  density is just 
$$
\mathcal{H}^{(1)}+\mathcal{H}^{(2)}=8M^2\left[\psi(x^1,x^2)
  +k^2\right]\frac{|u|^2}{(1+|u|^2)^2} 
$$
with
\begin{eqnarray}
\psi(x^1,x^2)\equiv
\frac{n^2}{|w_{\delta}|^2}+\frac{m^2}{|w_{-\delta}|^2}+
nm\frac{\bar{w}_{\delta}w_{-\delta}+
w_{\delta}\bar{w}_{-\delta}}{|w_{\delta}|^2|w_{-\delta}|^2}.\nonumber  
\end{eqnarray}
 Two examples of integrands $\frac{|u|\psi(x^1,x^2)^2}{(1+|u|^2)^2}$
 and $\frac{|u|}{(1+|u|^2)^2}$ are sketched in 
 Figs. \ref{pos}  and \ref{neg}. They correspond to the cases  
of $n=3$ and $m=2$ (Fig. \ref{pos}) and $m=-1$ (Fig. \ref{neg}).

If we interpret the solution $u(z,y_+)$ as describing a vortex we see
that $\mathcal{H}$  
describes its energy density per unit length of the vortex. Then this 
energy per unit length is given by a two dimensional  integral (with
integration over  the coordinates $x^1$ and $x^2$) and is given by 
\begin{eqnarray}
\mathcal{E}=8\pi M^2\left[K_{(n,m)}+k^2I_{(n,m)}\right],
\end{eqnarray}
where 
\begin{eqnarray}
\label{posia}
K_{(n,m)}\equiv
\frac{1}{\pi}\int_{R^2}dx^1\,dx^2\,\psi(x^1,x^2)\frac{|u|^2}{(1+|u|^2)^2} 
\end{eqnarray}
and
\begin{eqnarray}
\label{posi}
I_{(n,m)}\equiv\frac{1}{\pi}\int_{R^2}dx^1\,dx^2\,\frac{|u|^2}{(1+|u|^2)^2}.
\end{eqnarray}
So, according to (\ref{energytopnoether}), $K_{(n,m)}$ is related to
the topological charge and $I_{(n,m)}$ to the Noether charge. 

Note that in this vortex interpretation $K_{(n,m)}$ is the total energy of
the solution (\ref{one}) 
with $k=0$, which is then also a solution of the ($2+0$) dimensional $CP^1$ model.
This solution is purely topological and so the integrand
$\mathcal{H}^{(1)}$ is a total derivative, as is clear from
(\ref{niceidentity}),   
and moreover, is also the density of the topological charge.
Hence, this integrand tells us where the vortices are positioned (thus
for the configuration 
in Fig. \ref{pos} one vortex is located at $(-1.3,0)$ and two
vortices are on top of
each other at $(1.3,\,0)$). 
Similarily, the vortices in Fig. \ref{neg} are located around (5.0,\,0)).

The integral $K_{(n,m)}$ describes the total (2+0) dimensional energy
of these topological solitons (which is also proportional to their
topological charge) and so is independent of the parameter $\delta$
which characterizes the ``distance'' between the two vortices. As is well
known, and it is easy to check,  $K_{(n,m)}(\delta)$ depends on $n$ and
$m$ in the following way: 
\begin{enumerate}
\item
when $nm>0$ then $K_{(n,m)}(\delta)=|n+m|$ for any $\delta$,
\item
when $nm<0$ then $K_{(n,m)}(\delta\neq0)=\max(|n|,|m|)$,
otherwise one gets $K_{(n,m)}(\delta=0)=|n+m|$. 
\end{enumerate}

\begin{figure}
\begin{center}
\includegraphics[width=0.45\textwidth]{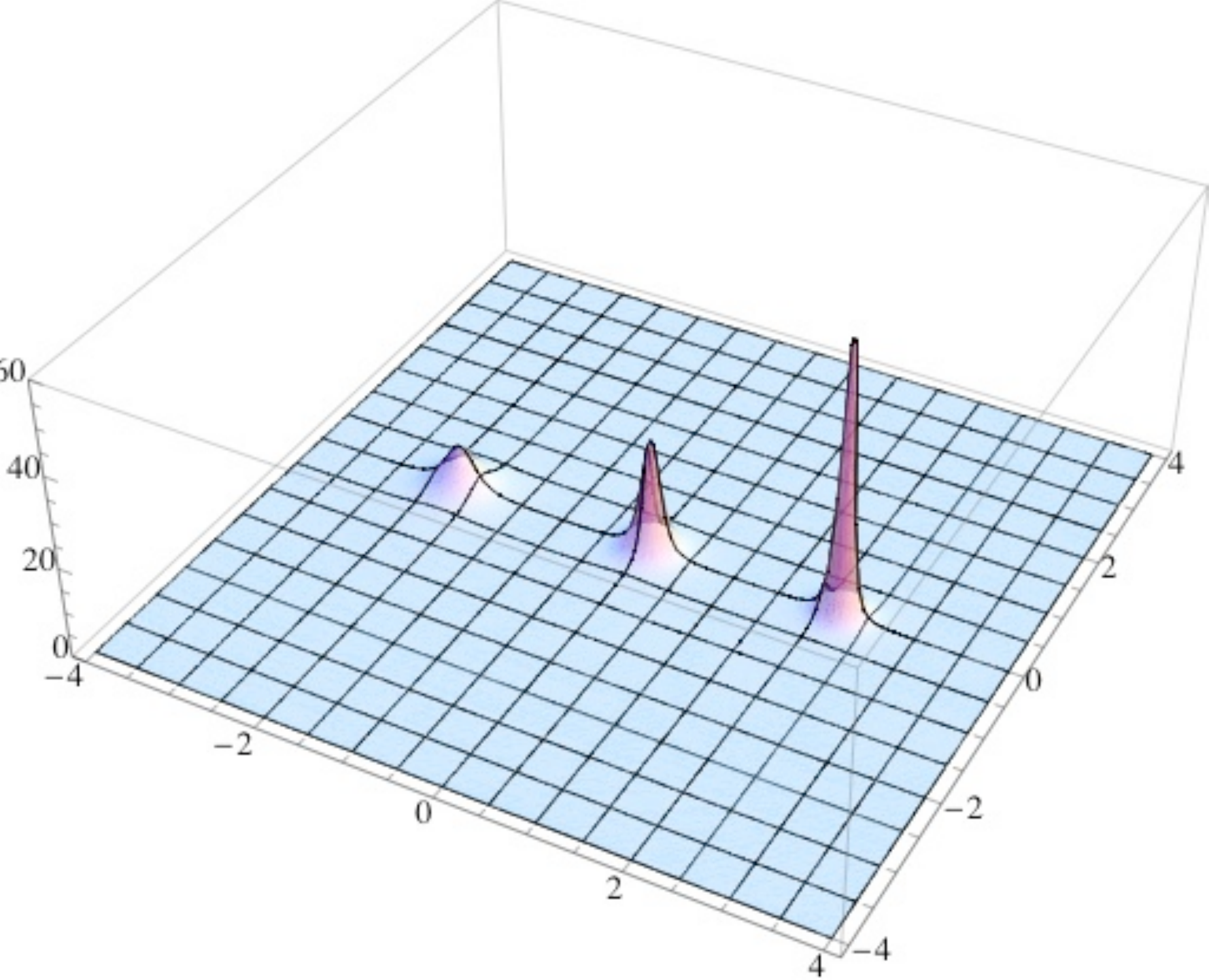}
\includegraphics[width=0.45\textwidth]{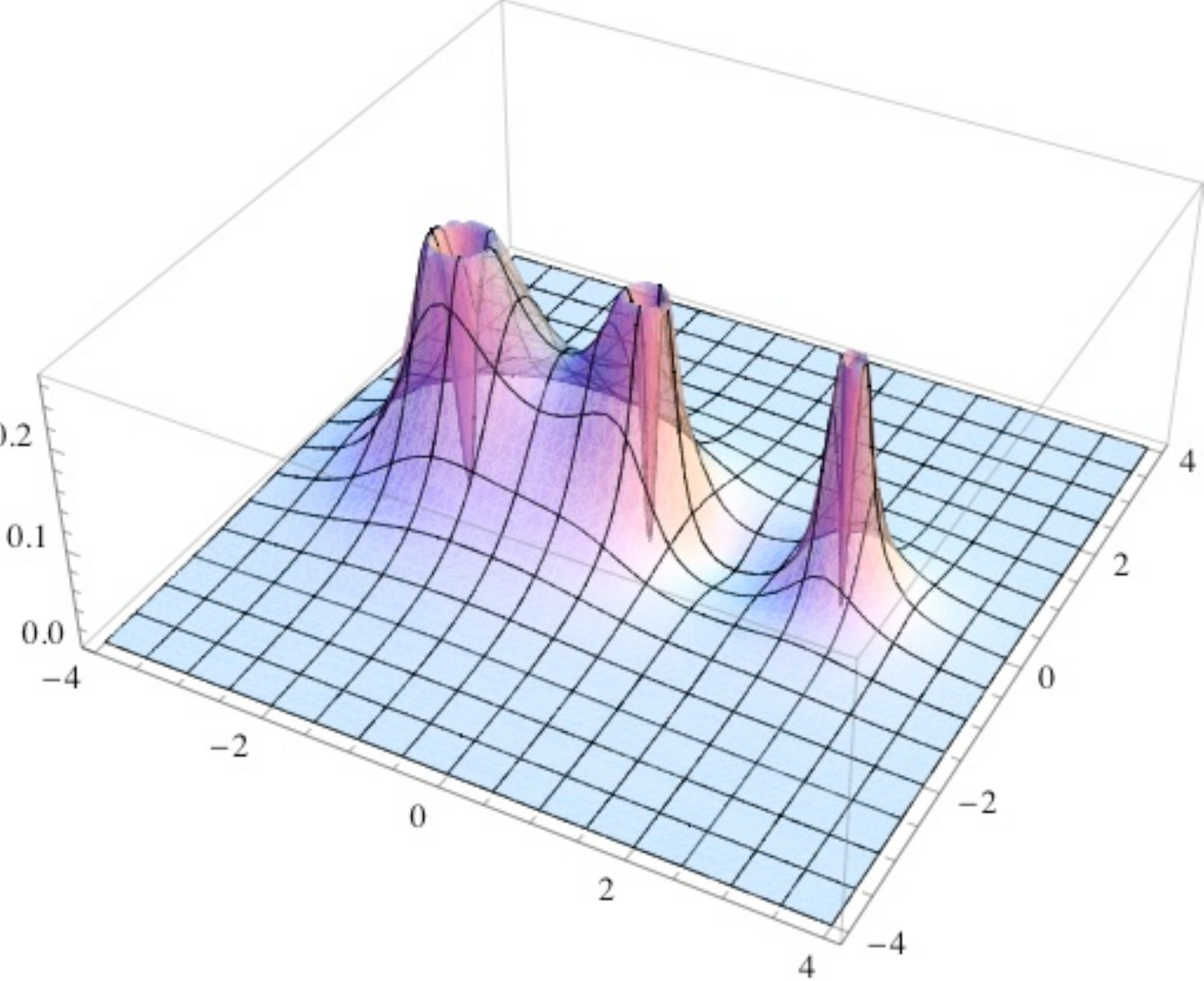}
\caption{The expressions $\frac{|u|^2\psi(x^1,x^2)}{(1+|u|^2)^2}$ (on
  left) and $\frac{|u|^2}{(1+|u|^2)^2}$ (on right) for the solution
  $u(z,y_+)=\frac{z(z^2-a^2)}{z-b}e^{iky_+}$ with $a=2.2$ and
  $b=1.0$.}
\label{map}
\end{center}
\end{figure}

\begin{figure}
\begin{center}
\includegraphics[width=0.45\textwidth]{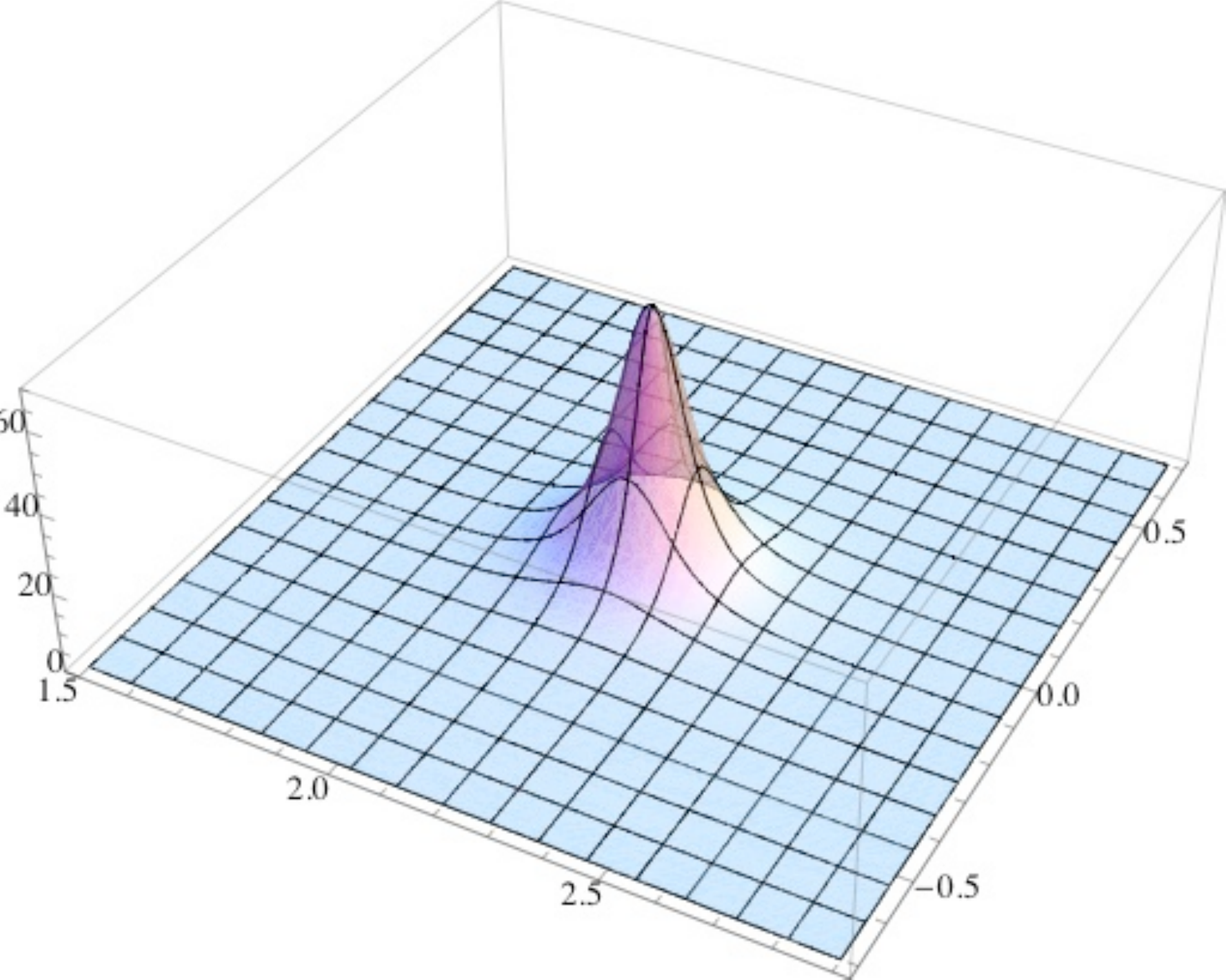}
\includegraphics[width=0.45\textwidth]{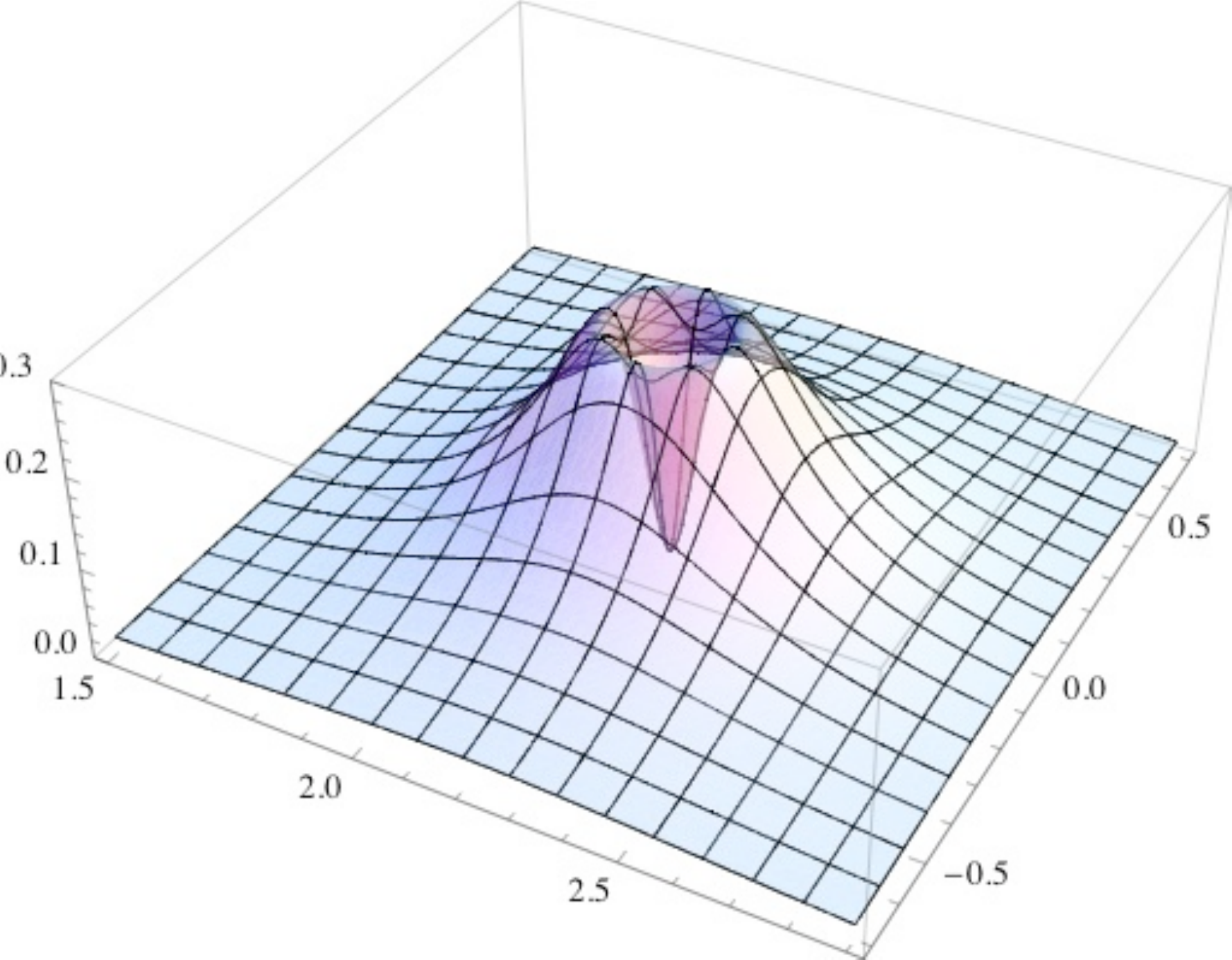}
\caption{A magnification of some region of Fig. \ref{map} where the most to right peak is located. The function $\frac{\psi |u|^2}{(1+|u|^2)^2}$
  (on left) has a local maximum at the point (2.2, 0) whereas the
  function $\frac{|u|^2}{(1+|u|^2)^2}$ 
(on right) takes value zero which is simultaneously its local  
  minimum.
\label{pick}} 
\end{center}
\end{figure}

In Fig. \ref{map} we present similar densities for a field conguration (also a
solution of the equations of motion)  
of the form
\begin{equation}
u(z,y_+)\,=\,e^{iky_+}\frac{z(z^2-a^2)}{z-b}
\end{equation}
for $a=(2.2,0.0)$ and $b=(1.0,0.0)$. We see clearly 3 peaks in each
picture, although each peak  on the right of the picture
corresponds to a peak with a `crater' at the centre of it. In Fig.
\ref{pick} we present  magnified pictures  
of the most to the right peaks of Fig. \ref{map}. This shows this `crater'
very clearly. As the first term describes the topological charge 
density we note that the energy density of the topological charge of
each vortex has a maximum at the position of each vortex  
and then decreases as you move away from this position.
The density of the other charge (or charges for $CP^N$ with $N>1$),
which are described by the other term is also located 
roughly at the positions of the vortices. However, the crater
like-structure, shows that at the exact positions 
of the vortices it is zero, then increases to a maximum in a small
circle around the exact position  
of the vortex and finally falls to zero as we move away from the vortices.

All this can be explained by the form of the functions in
(\ref{posia}) and (\ref{posi}), especially in the limit of well
separated vortices. 
In this case the positions are given by the zeros of $u$. As $\psi
\vert u\vert^2$ goes to a constant when $u\rightarrow 0$ 
the positions correspond to the maxima of the topological charge
density (see (\ref{posia})). For the Noether 
charge densities the situation is different.
The last term of the expression  given in (\ref{posi}) does not have
the $\psi$ factor and  
hence the integrand of (\ref{posi}) vanishes at this point. The
maximum is nearby and, when $u\rightarrow\infty$,  
the integrand, again, goes to zero. Thus the two charges are located
roughly at the same places but their charge distributions 
are slightly different; for the topological charge we have a simple
peak while for the Noether charge a `crater' like structure.

As we have already said before, taking into consideration only the first term $K_{(n,m)}$ one can
already see that the total energy of our solutions is infinite (as the
integrand does not depend  
on $x^3$ we clearly have a linear divergence). This is, of course,
consistent with our  
interpretation of the solutions are describing vortices, {\it i.e.}
structures whose energy is infinite but  
  the energy density per unit length is finite.

 However, the contribution to the total energy that comes from the
 second term, proportional to $k^2$, is also infinite.  
Bearing in mind our vortex interpretation of the solution the
behaviour of $I_{(n,m)}$ is worth studying in more detail. First, we
observe that even the integral $I_{(n,m)}$  can be divergent; in fact,
it is easy to check that $I_{(n,m)}$ diverges when the pairs $(n,\,m)$
satisfy  
the inequality $|n+m|<2$. Hence, the vortex-like interpretation has to be
reconsidered in a new light. 
Some vortex-like configurations appear to have  infinite energy per
unit length. 
However, looking at this problem in more detail we note something even
more interesting -  
the integral $I_{(n,m)}(\delta)$ can depend on $\delta$ in a nontrivial way.

\begin{figure}
\begin{center}
\includegraphics[width=0.65\textwidth]{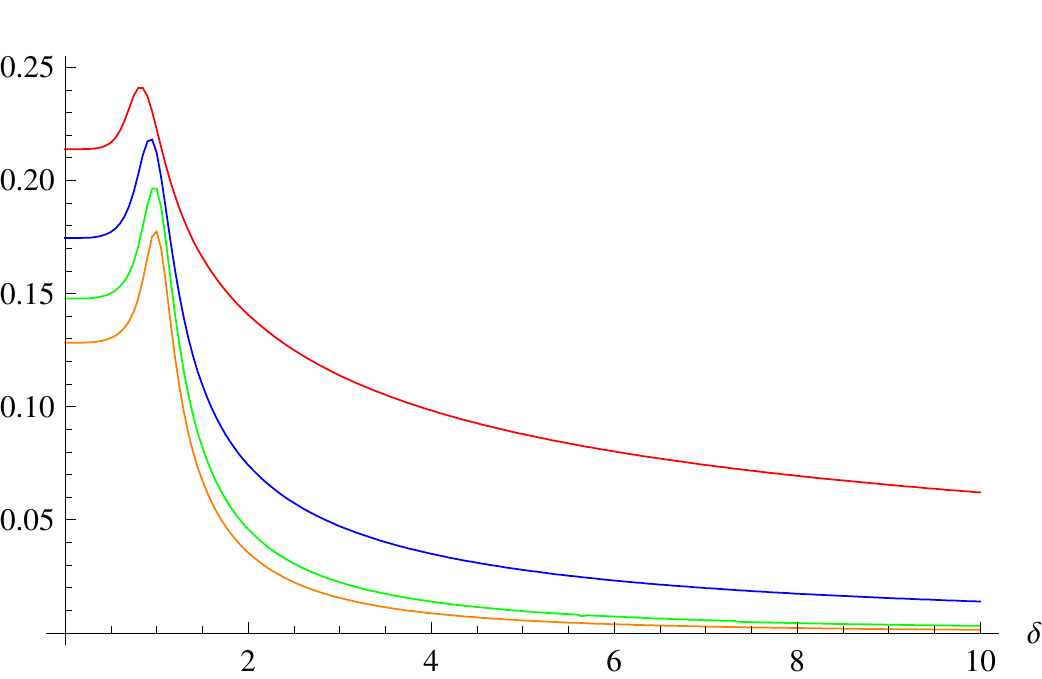}
\caption{The integrals (from the top) $I_{(4,1)}(\delta)$, $I_{(4,2)}(\delta)$, $I_{(4,3)}(\delta)$, $I_{(4,4)}(\delta)$. }
\label{pos2}
\end{center}
\end{figure}

\begin{figure}
\begin{center}
\includegraphics[width=0.65\textwidth]{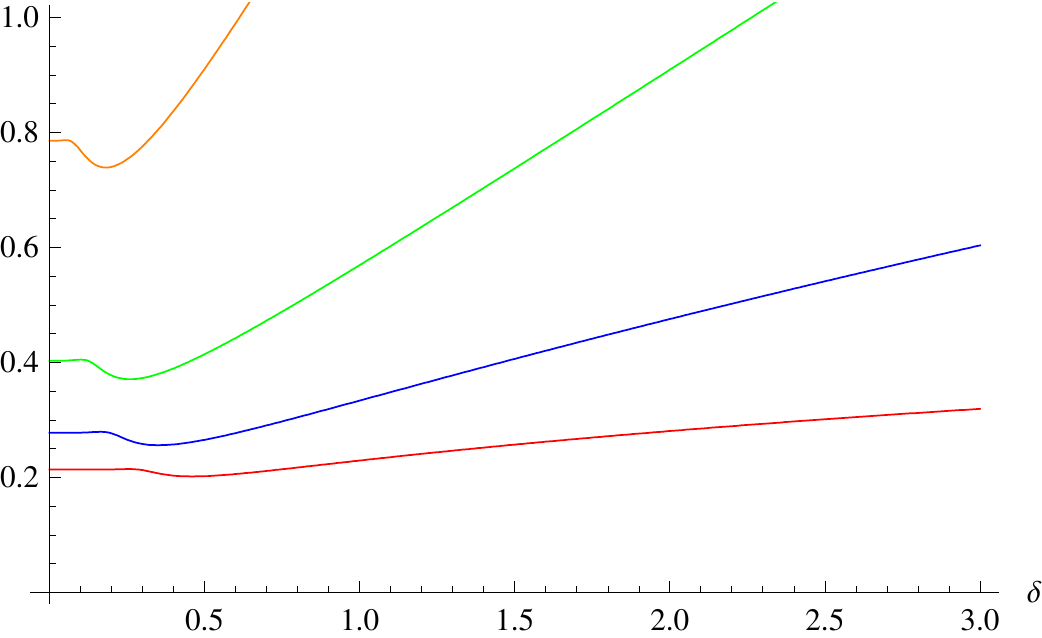}
\caption{The integrals (from the bottom) $I_{(6,-1)}(\delta)$, $I_{(6,-2)}(\delta)$, $I_{(6,-3)}(\delta)$, $I_{(6,-4)}(\delta)$.}
\label{neg2}
\end{center}
\end{figure}

To discuss this we consider the pairs $(n,\, m)$ which satisfy $|n+m|\ge 2$ 
with both $n$ and $m$ $\neq0$. Note that our solutions possess  two qualitatively
different asymptotic behaviours of $I_{(n,m)}(\delta)$ for large
$\delta$. Namely, when $nm>0$ the integral tends to zero as
$\delta\rightarrow\infty$, (see Fig. \ref{pos2}), and  when $nm<0$, in the
same limit, it diverges  
  (see Fig. \ref{neg2}). Moreover, in the last case the integral
 $I_{(n,m)}(\delta)$ has a local minimum for some finite $\delta$

\begin{figure}
\begin{center}
\includegraphics[width=0.65\textwidth]{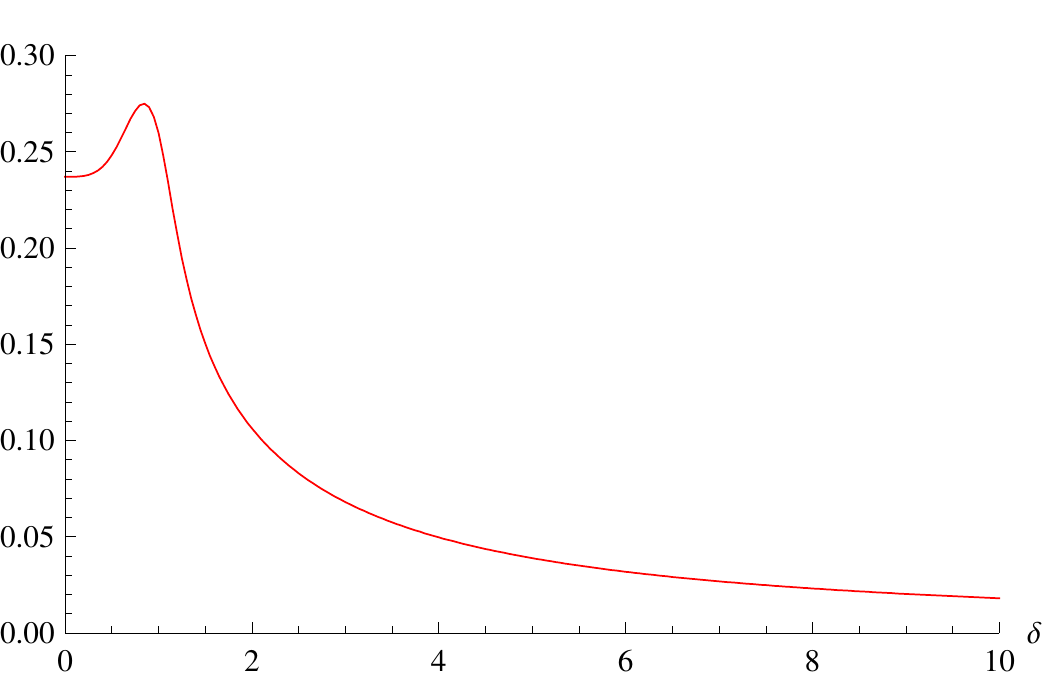}
\caption{The integral $I(\delta)\equiv\frac{1}{8\pi M^2}\int_{R^2}dx^1dx^2\mathcal{H}^{(2)}$ for the $CP^2$ solution $u_1(z,y_+)=(z-\delta)^{-1}(z+\delta)^{-1}e^{-iy_+}$ and $u_2(z,y_+)=(z-\delta)^{2}(z+\delta)^4e^{iy_+}$.}
\label{cp2_2}
\end{center}
\end{figure}

\begin{figure}
\begin{center}
\includegraphics[width=0.65\textwidth]{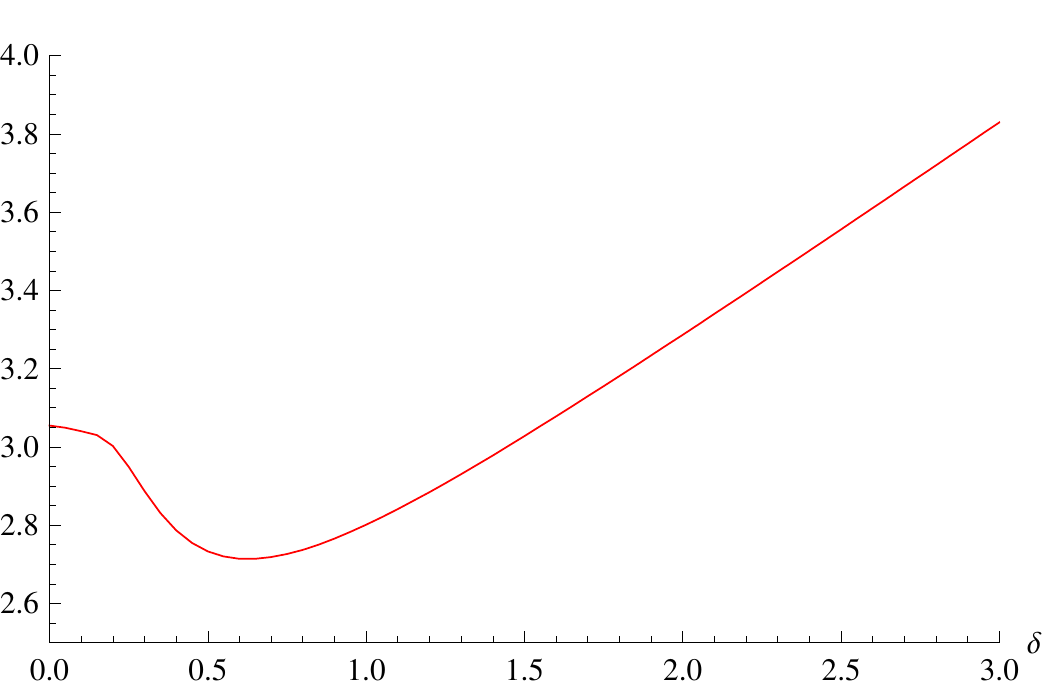}
\caption{The integral $I(\delta)\equiv\frac{1}{8\pi M^2}\int_{R^2}dx^1dx^2\mathcal{H}^{(2)}$ for the $CP^2$ solution $u_1(z,y_+)=(z-\delta)^{-1}(z+\delta)e^{-iy_+}$ and $u_2(z,y_+)=(z-\delta)^{-2}(z+\delta)^4e^{iy_+}$.}
\label{cp2_1}
\end{center}
\end{figure}

 The $CP^2$ case is qualitatively very similar to the $CP^1$ one. As an illustration 
in Figs. \ref{cp2_2} and \ref{cp2_1} we present ther plots of the energy per unit length for the $CP^2$ solutions.

At first sight this looks like a paradox, since one would naively expect that
the dependence of  
the energy on the distance $\delta$ would lead to a force between the
vortices. Therefore, only the configurations corresponding to the
minimum of the energy would be expected to correspond to true
solutions of the equations of motion.
The resolution of this `paradox' is based on the observation that  the
total energy of all these solutions is infinite. The solutions  
of the equations of motion minimize the total energy but here the
total energy is infinite. 
It does not matter that the value of the energy per unit length
depends on $\delta$ and is finite - the total energy is still
infinite. Hence we are 
in the situation in which the vortices can have different energies per
unit length and still be at rest with respect to each other. Another
way of phrasing the resolution of the paradox is that even though
there is a finite force between the vortices (due to the dependence of
the energy per unit of length on the distance $\delta$) their masses
are infinite and so there is no acceleration. 

The dependence of the energy per unit of length on the distance
$\delta$ raises some possibilities for further investigations. Notice
that $\delta$ is a  distance in the $x^1\,x^2$ plane, and the
dependence of the energy on $\delta$ occurs only due to the term
depending on the two other dimensions, namely $x^3$ and $x^0$.
Suppose now that one
considers the same model but in a four dimensional space that contains
the ($2+1$) Minkowski space as a border (for instance AdS$_4$). It
would be interesting to verify if the same effect takes place here too. If it
does, then one could interpret the solution as $CP^N$ lumps on a
($2+1$) Minkowski space (the border) which acquire an interaction due
to the part of the Hamiltonian which lives in the bulk of the four
dimensional space. That would be a novel feature which certainly could have
many interesting applications.    
  
\section{Some Conclusions}

In this paper we have demonstrated that the $CP^N$ model in (3+1)
dimensions has many classical solutions. The ones we have discussed
here correspond to field configurations described by functions of two
variables ($x^1+i\epsilon_1 x^2$ and $x^3+\varepsilon_2\, x_0$). As the energy of
these configurations is infinite (as the energy density is independent
of $x^3$) these solutions describe vortices  
with the time dependence  corresponding to the rotation of the vortices
in the internal space 
(in this property they resemble a little `Q-lumps'\cite{Leese}.) We
have shown that they possess many interesting properties (like
interesting distributions of topological charge and of the Noether  
charge). One of the more unusual properties is their dependence on the
distance between  
the lines of charge: the energy density (of energy per unit length) of
one vortex is infinite but of two vortices can depend on the distance
between them and possess a minimum at a specific 
value of this distance. This suggests that vortices which are located
at non-minimal distances 
may be unstable and so could reduce their energy per unit length by moving
towards this optimal  
configurations. However, their configurations are solutions for any
distance as their infinite  
`inertia' stops them from moving towards  them without an external push.

We are now looking at other properties of these and other solutions.

{\bf Acknowlegment:} L.A. Ferreira and W.J. Zakrzewski would like to thank
the Royal Society (UK)
for a grant that helped them in carrying out this work. L.A. Ferreira
is partially supported by CNPq (Brazil) and P. Klimas is supported by
FAPESP (Brazil).

\end{document}